\documentclass[conference,letterpaper]{IEEEtran}
\IEEEoverridecommandlockouts
\usepackage{amsmath,amssymb,amsfonts}
\usepackage{hyperref}
\usepackage{graphicx}
\usepackage{bbm}
\graphicspath{{./figs/}}

\def\wt{\mathop{\mathrm{wt}}}    
\def\tr{\mathop{\mathrm{Tr}}}
\def\ket#1{\left|#1\right\rangle}
 \newcommand{\ba}{{\bf a}}\newcommand{\bA}{{\bf A}}
 \newcommand{\bb}{{\bf b}}

 \newcommand{\e}{{\bf e}}
 \newcommand{\f}{{\bf f}}\newcommand{\F}{{\bf F}}
\newcommand{\bg}{{\bf g}}\newcommand{\bG}{{\bf G}}
 \newcommand{\bH}{{\bf H}}
 \newcommand{\bI}{{\bf I}}

 \newcommand{\bM}{{\bf M}}

 \newcommand{\bQ}{{\bf Q}}
 
 \newcommand{\bs}{{\bf s}}
 
 \newcommand{\bu}{{\bf u}}
 \newcommand{\bv}{{\bf v}}
 
 \newcommand{\bx}{{\bf x}} 
 \newcommand{\x}{{\bf x}} 
 \newcommand{\by}{{\bf y}}

\begin{document}

\title{Quantum convolutional data-syndrome codes
}

\author{\IEEEauthorblockN{%
    Weilei Zeng\IEEEauthorrefmark{1}$^a$, 
    Alexei
    Ashikhmin\IEEEauthorrefmark{2}, %
    Michael Woolls\IEEEauthorrefmark{1}$^b$, and %
    Leonid P. Pryadko\IEEEauthorrefmark{1}$^c$\thanks{This work was
      supported by the NSF Division of Physics grant \#1820939.}}
  \IEEEauthorblockA{\IEEEauthorrefmark{1}\textit{Department of Physics \&
      Astronomy} \textit{University of California} \textit{Riverside,
    California, USA}}
  \IEEEauthorblockA{$^a$wzeng002@ucr.edu\quad
    $^b$mwool001@ucr.edu\quad
    $^c$leonid.pryadko@ucr.edu}
  \IEEEauthorblockA{\IEEEauthorrefmark{2}\textit{Bell Labs, Nokia} 
    \textit{Murray Hill, NJ, USA}\quad alexei.ashikhmin@nokia.com}}

\maketitle

\begin{abstract}
  We consider performance of a simple quantum convolutional code in a
  fault-tolerant regime using several syndrome measurement/decoding
  strategies and three different error models, including the circuit
  model.
\end{abstract}


\section{Introduction}
Quantum stabilizer codes are designed to be robust against qubit
errors.  However, syndrome measurement cannot be done perfectly:
necessarily, there are some measurement errors whose probability grows
with the weight of the checks (stabilizer generators).  Furthermore, both the
syndrome measurement protocol and the syndrome-based decoding have to
operate in a fault-tolerant (FT) regime, to be robust against errors
that happen during the measurement.

When all checks have relatively small weights, as in
the case of the surface codes, one simple approach is to repeat
syndrome measurement several
times\cite{Dennis-Kitaev-Landahl-Preskill-2002}.  Then, FT
syndrome-based decoding can be done in the assumption that the data
errors accumulate while measurement errors be independently
distributed.  While there is always a non-vanishing probability to
have some errors at the end of the cycle, what matters in practice is
the ability to backtrack all errors after completion of several rounds
of measurement.

Another approach it to measure an overcomplete set of stabilizer
generators, using redundancy to recover the correct syndrome.  Such an
approach was used in the context of higher-dimensional
toric and/or color
codes\cite{Bombin-Chhajlany-Horodecki-MartinDelgado-2013,%
  Duivenvoorden-Breuckmann-Terhal-2017}, the data-syndrome (DS)
codes\cite{Fujiwara-2014,%
  Ashikhmin-Lai-Brun-2014,Ashikhmin-Lai-Brun-2016}, and single-shot
measurement
protocols\cite{Bombin-2015,Brown-Nickerson-Browne-2016,Campbell-2018}.
Here decoding is done in the assumption that data error remains the
same during the measurement.

We note that with both approaches, the error models assumed for
decoding do not exactly match the actual error probability
distribution.  In particular, any correlations between errors in
different locations and/or different syndrome bits are typically
ignored.  Nevertheless, simulations with circuit-based error models
which reproduce at least some of the actual correlations show that
both the repeated syndrome measurement
protocol\cite{Wang-Fowler-Hollenberg-2011,Landahl-2011} and the
syndrome measurement protocols relying on an overcomplete set of
generators\cite{Duivenvoorden-Breuckmann-Terhal-2017} can result in
competitive values of FT threshold.

The choice of the measurement protocol is typically dictated by the
structure of the code, specifically, availability of an overcomplete
set of stabilizer generators of the minimum weight.  Such an approach
is expected to be practical when typical gate infidelities are
comparable with the probability of an incorrect qubit measurement.
However, there is also a price to pay: codes with redundant sets of
small-weight checks can be generally expected to have worse
parameters.

On the other hand, if the physical one- and two-qubit gates are
relatively accurate, it may turn out more practical to measure
redundant sets of checks which include stabilizer generators of higher
weights.  Then, a DS code can be designed from any stabilizer
code\cite{Fujiwara-2014,%
  Ashikhmin-Lai-Brun-2014,Ashikhmin-Lai-Brun-2016}.  As a result, one
faces a problem of constructing an optimal measurement protocol given
the known gate fidelities and measurement errors.

In this work we compare several single-shot and repeated
measurement/decoding protocols for a simple quantum convolutional
code\cite{Forney-Grassl-Guha-2007} with the parameters $[[24,6,3]]$
and syndrome generators of weight 6.  We construct several
computationally efficient schemes using the classical Viterbi
algorithm\cite{Viterbi-1967,johannesson2015fundamentals} to decode
data and syndrome errors sequentially or simultaneously, and compare
their effectiveness both with phenomenological and circuit-based
depolarizing error models.  In particular, we show that a DS code
which requires measuring checks of weight up to $w_\mathrm{max}=9$ has
performance (successful decoding probability) exceeding that of the
repeated measurement scheme when single-qubit measurement error
probability $q_1$ equals ten times the gate error probability $p_1$
(taken to be the same for Hadamard and CNOT gates).

\section{Background}
\label{sec:background}
Let
$\mathcal{P}_n=\{c\, M_1\otimes \cdots \otimes M_n:M_j\in \{I, X,
Y, Z\} \} $, with phase $c\in \{\pm 1,\pm i\}$ be the $n$-qubit Pauli group;
elements with $c=\pm1$ have eigenvalues $\pm 1$.  Any
$G= c\, M_1\otimes \cdots \otimes M_n\in \mathcal{P}_n$ can be
represented, up to a phase, by length-$n$ vector
$\bg=(g_1,\ldots,g_n)\in
\mathbb{F}_4^n,~\mathbb{F}_4=\{0,1,\omega,\bar{\omega}\}$, where 
$\omega^2\equiv \bar\omega=\omega+1$, and
$g_j=0,1,\omega,\bar{\omega}$, if $M_j$ is $I$, $X$, $Z$, or $Y$,
respectively.  The weight $\wt(\bg)$ of $\bg$ is the number of its
nonzero elements $g_j\not=0$.  A product of two Pauli operators $X$
and $Y$ is mapped into a sum of the corresponding vectors, $\bx +\by$.
Further, a pair of Pauli operators commute iff the trace inner
product,
\begin{equation}
  \label{eq:trace-product}
  \bx *\by^T\equiv \sum\nolimits_{i=1}^n\tr(x_i\bar{y}_i),
\end{equation}
of the corresponding vectors is zero, $\bx*\by^T=0$.  Here
$\tr(x)=x+x^2$ is the trace map from $\mathbb{F}_4$ into
$\mathbb{F}_2$, and $\bar y$ is the conjugation of $y\in \mathbb{F}_4$
which interchanges $\omega$ and $\bar\omega$.

An $[[n,k]]$ quantum stabilizer code ${\cal Q}$, a subspace of the
$n$-qubit Hilbert space, encodes $k$ qubits into $n$, and its rate is
defined as $R=k/n$.  Any code vector (quantum state) $\ket\psi\in
{\cal Q}$
is stabilized by a \emph{stabilizer} group ${\cal S}$, an  Abelian
subgroup of $\mathcal{P}_n$ such that $-I^{\otimes n}\not\in{\cal S}$.
Such a subgroup is mapped to an \emph{additive} code ${\cal C}$ of
size $|{\cal C}|=2^{n-k}$ over $\mathbb{F}_4$, specified by the
generator matrix $\bG$ whose rows $\bg_j$ correspond to the Pauli
generators $G_j\in{\cal S} $.  The additivity means that for any
$\bx, \by\in {\cal C}$ we have $\bx+\by\in {\cal C}$.  A detectable
error $E\in {\cal P}_n$ has non-zero commutators with one or more
generators of ${\cal S}$; the corresponding vector
$\e\in\mathbb{F}_4^n$ has a non-zero syndrome $\bs^T=\bG*\e^T$.
Vectors corresponding to undetectable errors form the dual (with
respect to the trace inner product) code ${\cal C}^\perp$ of size
$|{\cal C}^\perp|=2^{n+k}$.  Since ${\cal S}$ is Abelian, one
necessarily has ${\cal C}\subseteq {\cal C}^\perp$, and for this
reason ${\cal C}$ is called {\em self-orthogonal}.  Elements of
${\cal S}$ act trivially on the code ${\cal Q}$; thus the distance $d$
of a quantum code is defined as the minimum weight of an element of
${\cal C}^\perp\setminus {\cal C}$ \cite{Calderbank-1997}.

For numerics in this work we use the family of quantum convolutional
codes (QCCs) of length $3(k+2)$, $k=1,2,\ldots$, based on linear
$(3,1)$ self-orthogonal convolutional codes whose generator matrces
are constructed\cite{Forney-Grassl-Guha-2007} by $k+1$ shifts of the
row $\bg_1=(111|1\omega\bar\omega)$.  The actual generating matrix of
the QCC ${\cal Q}_k$ with parameters $[[3(k+2),k,3]]$ is obtained by
adding a copy of the same rows multiplied by $\omega$, and four
additional rows for proper termination.  In the case $k=1$, the
stabilizer generating matrix has the form
{\small\begin{equation}
\bG ({\cal Q}_1)=\left(
  \begin{array}{ccc|ccc|ccc}
    1 &\omega & \overline \omega& & & \\
    \overline \omega & \omega & 1& & & \\
    1 & 1 & 1 &1 &\omega & \overline \omega\\
    \omega & \omega &\omega & \omega &\overline \omega &1 \\
      &&&   1 & 1 & 1 &1 &\omega & \overline \omega \\
      &&&     \omega & \omega &\omega & \omega &\overline \omega &1 \\
      &&&&&&    \overline \omega & \omega & 1\\
      &&&&&& 1 &1 &1\\
  \end{array}\right).\label{eq:QCC-G0}
\end{equation}}%

\section{Error models and data-syndrome codes}
\label{sec:error-model}
Unlike with classical codes, extracting a syndrome for a quantum code
involves a complicated quantum measurement which itself is prone to
errors.  To extract a syndrome bit corresponding to a row $\bg $ of
$\bG$, one must execute a unitary which involves a non-trivial
interaction [some single-qubit gate(s) and an entangling gate, e.g., a
quantum \texttt{CNOT}] with each of the $w\equiv \wt(\bg)$ qubits in
the support of $\bg$, then do a quantum measurement of one or more
auxiliary \emph{ancilla} qubit(s).  Data errors and measurement
(ancilla) errors can happen at every step of the process; moreover,
errors can propagate through measurement circuit unless it is designed
using FT gadgets to prevent error multiplication\cite{Shor-FT-1996}.
Error propagation can be simulated efficiently for any circuit
constructed from Clifford gates which map the Pauli group onto itself,
which is sufficient to simulate the performance of any stabilizer
code\cite{Gottesman-Kn-1998}.  In this work we simulated such a
\emph{circuit-based error} model (C), using depolarizing noise with
probability $p_1$ (randomly chosen $X$, $Y$, or $Z$ on every qubit in
the interval between subsequent gates, including null gates for idle
qubits), and additional ancilla measurement error with probability
$q_1$ \cite{Wang-Fowler-Hollenberg-2011,Landahl-2011}.

While in principle it is possible to account for all correlations
between the errors that may result from error propagation in a given
circuit, and design a corresponding decoder, 
it would be a daunting task.  Instead, one usually uses a decoder
designed for some phenomenological error model, and uses circuit model
(C) only to check the performance of such a decoder numerically.  We
consider two such error models.

Model (A) is a channel model where qubit errors (depolarizing noise
with probability $p$) happen before the measurement, while each
stabilizer generator (syndrome bit) is measured with independent error
probability $q$.  This
model\cite{Ashikhmin-Lai-Brun-2014,Ashikhmin-Lai-Brun-2016} is an
idealization of a situation where gate errors are small compared to
qubit preparation and measurement errors.  Clearly, model (A) can get
unphysical, as here one can extract the syndrome perfectly with
sufficient measurement redundancy.

This drawback is compensated somewhat in the phenomenological error
model (B) which includes several rounds of syndrome measurement, and
includes qubit errors that happen before each round (depolarizing
noise with probability $p$; these errors accumulate between
measurement rounds), and independent syndrome measurement errors with
probability $q$.  Both in the phenomenological model (B) and in the
circuit model (C) some errors may remain after the last round of error
correction; for simulations one includes an additional round with
perfect syndrome measurement\cite{Wang-Fowler-Hollenberg-2011}.

Phenomenological error models (A) and (B) can be used to construct DS
codes dealing both with qubit (data) and syndrome errors.  We start
with an $r\times n$ stabilizer generator matrix $\bG$, which may
include additional linearly-dependent rows, thus $r\ge n-k$.  In model
(A), we have a qubit error vector $\e\in\mathbb{F}_2^n$, and a
syndrome measurement error $\boldsymbol\epsilon\in\mathbb{F}_2^r$; the
extracted syndrome vector is given by
$\bs^T=\bG*\e^T+\boldsymbol\epsilon^T$.  To characterize DS codes, it
is convenient to consider mixed-field vector spaces, with elements
$\left(\e\,|\,\boldsymbol\epsilon\right)$, a pair of a quaternary and
a binary vectors.  For such pairs we define the inner product
\begin{equation}
  \label{eq:DS-inner}
  (\e_1,\boldsymbol\epsilon_1)\star
  (\e_2,\boldsymbol\epsilon_2)^T\equiv
  \e_1*\e_2^T+\boldsymbol\epsilon_1 \boldsymbol\epsilon_2^T.
\end{equation}
By analogy with stabilizer codes, we define an additive code
${\cal C}_{\rm DS}\subseteq\mathbb{F}_4^n\oplus \mathbb{F}_2^r$ with
the generator matrix
\begin{equation}
  \label{eq:G-DS}
  \bG_{\rm DS}=\left(\begin{array}[c]{c|c}\bG&\bI\end{array}\right), 
\end{equation}
and its dual with respect to the product (\ref{eq:DS-inner}),
${\cal C}_{\rm DS}^\perp$.  The two orthogonal DS codes satisfy
$|{\cal C}_{\rm DS}|\, |{\cal C}_{\rm DS}^\perp|=2^{2n+r}$.  Because
the original code ${\cal C}$ is self orthogonal, $\bG*\bG^T=0$, the
code ${\cal C}_{\rm DS}^\perp$ includes vectors in the form
$(\e\,|\,{\bf 0})$, where $\e=\boldsymbol\alpha \bG$ is an additive
combination of the rows of $\bG$,
$\boldsymbol\alpha\in\mathbb{F}_2^r$. The distance $d_{\rm DS}$ of
thus defined DS code is the minimum weight of a vector in
${\cal C}_{\rm DS}^\perp\setminus ({\cal C}\oplus {\bf 0})$, it is upper
bounded by the distance of the original quantum code ${\cal Q}$,
$d_{\rm DS}\le d$.

In phenomenological error model (B), with $\ell$-times repeated
syndrome measurement (including the final perfect measurement), we
denote qubit errors that occur before the measurement $t$ as
$\e_t\in\mathbb{F}_4^n$, and the corresponding measurement errors as
$\boldsymbol\epsilon_t\in\mathbb{F}_2^r$.  The qubit errors
accumulate, thus we can write for the syndrome $\bs_t$ obtained in the
$t$\,th round of measurement:
\begin{eqnarray*}
  \bG*(\e_1+\e_2+\ldots +\e_{t})+\boldsymbol\epsilon_{t}&=&\bs_{t}.
\end{eqnarray*}
In this work we do not attempt  simultaneous decoding of data and
syndrome errors over several rounds of measurement.  Instead we decode
them sequentially, using the accumulated errors 
$\hat{\e}_1+\hat{\e}_2+\ldots+\hat{\e}_{t-1}$ extracted at previous
decoding rounds to offset the error at time $t$.

\section{Convolutional DS codes}

Now, given an $[[n,k]]$ quantum code ${\cal Q}$ with the
(full-row-rank) generating matrix $\bG({\cal Q})$ of size
$(n-k)\times n$, we introduce redundant measurements by adding some
linearly dependent rows.  Without limiting generality, a set of $r'$
additional rows $\F=\bA \bG({\cal Q})$ can be obtained by multiplying
the original generating matrix by an $r'\times (n-k)$ binary matrix
$\bA$, so that the generating matrix \ref{eq:G-DS} of the resulting DS
code has the form
\begin{equation}
  \label{eq:G-DS}
  \bG_{\rm DS}=\left(
    \begin{array}[c]{c|cc}
      \bG({\cal Q})& \bI_{n-k} &  \\
      \bA \bG({\cal Q}) & & \bI_{r'}
    \end{array}\right).
\end{equation}
This matrix has additive rank $r\equiv (n-k)+r'$ equal to the number of
rows.  It is convenient to rewrite this matrix in the following
row-equivalent form,
\begin{equation}
  \label{eq:G-DS-prime}
  \bG_{\rm DS}'=\left(
    \begin{array}[c]{c|cc}
      \bG({\cal Q})& \bI_{n-k} &  \\
                   &\bA & \bI_{r'}
    \end{array}\right).
\end{equation}
 Denote $[\bG({\cal Q})]^\perp$ the additive dual of
$\bG({\cal Q})$ with additive rank $n+k$, and $\bM$ a
matrix such that $\bG({\cal Q})\,\bM^T=\bI_{n-k}$.  It is then easy to
see that the matrix
\begin{equation}
  \label{eq:H-DS}
  \bH_{\rm DS}=\left(
    \begin{array}[c]{c|cc}
      [\bG({\cal Q})]^\perp&  &  \\
                 \bM  &\bI_{n-k} & \bA^T
    \end{array}\right).
\end{equation}
has additive rank $(n+k)+(n-k)=2n$, while
$\bG_{\rm DS}'\star \bH_{\rm DS}^T=0$. Thus, $\bH_{\rm DS}$ generates
the code ${\cal C}_{DS}^\perp$.

We can now discuss the choice of the matrix $\bA$.  First, we obtain
redundant syndrome bits by measuring operators $F_j$ corresponding to
the rows $\f_j$ of the matrix $\F$.  Since the corresponding error
grows with the operator weight, we want to choose matrix $\bA$ to
ensure that row weights of $\F$ be small.  
Second, we want to choose $\bA$ so that the binary linear code
generated by $(\bI_{n-k},\bA^T)$ has a large minimum distance.  Third
important issue is the decoding complexity.  Given the structure of
the matrix $\bH_{\rm DS}$, see Eq.\ (\ref{eq:H-DS}), it is natural to
choose $\bA^T$ to form a generator matrix of a classical convolutional
code.  Quantum DS codes (\ref{eq:G-DS}) obtained from a quantum
convolutional code ${\cal Q}$ with such an $\bA$ we call
\emph{convolutional DS codes}.

\section{Decoding of Convolutional DS Codes}
Big advantage of classical convolutional codes is that one can use the
maximum-likelihood Viterbi decoding using a code trellis
\cite{johannesson2015fundamentals}. The ``stripe'' form of a generator
matrix of a convolutional code (with small band width) ensures that
its code trellis has relatively small number of states, which means
that the Viterbi decoding has relatively small complexity.

In our case, it is not immediately obvious how to construct a
code trellis with a manageable number of states, since neither
$\bG_{\rm DS}$ nor $\bH_{\rm DS}$ has the ``stripe'' form. However, we
show that $\bG_{\rm DS}$ can be transformed into the stripe form.

Instead of presenting a general algorithm for this, we will consider a
small example. Let $\bG({\cal Q})$ and $\bA$ be generated by vectors
$(\bv_1|\bv_2|\bv_3)$ and $(\bu_1|\bu_2|\bu_3)$, respectively, and
assume that $\bv_i$ and $\bu_i$ have lengths $n$ and $n'$.  Then, in a
particular case, the DS code generator (\ref{eq:G-DS-prime}) has the
form {\small
  \begin{align*}
    &\bG_{\rm DS}'=\\
    &\left(\begin{array}{cccc ccc cccccc}
             \bv_1 &\bv_2 &\bv_3 & & & 1 & & & & & & & \\
                   & \bv_1 &\bv_2 &\bv_3 & & &  1 & & & & & & \\
                   & &\bv_1 &\bv_2 &\bv_3 & & &  1 & & & & &  \\
                   & & & & & \bu_1^T & & & \bI & & & & \\
                   & & & & & \bu_2^T & \bu_1^T & & & \bI  & & & \\
                   & & & & & \bu_3^T & \bu_2^T & \bu_1^T & & & \bI  & & \\
                   & & & & &        & \bu_3^T & \bu_2^T & & &  & \bI &  \\
                   & & & & &        &         & \bu_3^T & & &  & & \bI 
           \end{array}\right),
  \end{align*}%
}%
where $\bI$ is the $n'\times n'$ identity matrix.  With an appropriate
permutation of columns and rows, we can transform the above matrix
into the form {\small
  \begin{align*}
    &\bG_{\rm DS}''=\nonumber \\
    &\left( 
      \begin{array}{cccc|ccc|cccccc}
        \cline{5-7}
        \bv_1 & \bv_2 &   &   &\bv_3 & 1 & & \\
              &       & \bI &   &       & \bu^T_1& &  \\
              & \bv_1 &   &   & \bv_2 &         &   & \bv_3 & 1 \\
              &       &   & \bI &       & \bu^T_2 &   &       & \bu^T_1 \\
              &       &   &   & \bv_1 &         &   & \bv_2 &         &   & \bv_3 & 1\\
              &       &   &   &       & \bu^T_3 & \bI &       & \bu^T_2 &   &       & \bu^T_1 \\
             \cline{5-7}
              &       &   &   &       &         &   &       & \bu^T_3 & \bI &       & \bu^T_2 &   \\
              &       &   &   &       &         & &  &  & & &  \bu^T_3 & \bI
      \end{array} \right),
  \end{align*}}%
where we marked the small matrix block that defines the repeating
section of the syndrome trellis. 
Now, the method in Ref.~\cite{sidorenko1994decoding} gives  the
syndrome trellis, a particular form of the code trellis. 

Let $(\bv,\bs)\in C_{\rm DS}^\perp$ and define the received vectors
$\x=\bv+\e\in \mathbb{F}_4^n$,
$\by=\bs+\boldsymbol\epsilon\in \mathbb{F}_2^{n-k+r'}$, where $\e$ and
$\boldsymbol\epsilon$ are qubit and syndrome errors, respectively.
The syndrome allows us to efficiently conduct the Viterbi minimum
distance decoding (MDD) using $(\x,\by)$ as an input:
$$
\mathop{\rm MDD}(\x,\by)=\arg\min_{(\ba,\bb)\in C_{\rm DS}^\perp}
\wt(\ba-\x)+\wt(\by-\bb).
$$
However, unlike in the classical case where we receive $(\x,\by)$ from
a channel, in the quantum case we have only $\by=\bs+\e$, and we do
not have $\x$. 
It is easy to check that in this case the correct minimum distance
decoding corresponds to $\mathop{\rm MDD}({\bf 0},\by)$.  For
simulations in this work we implemented a version of Viterbi decoding
for non-binary classical codes with known symbol error probabilities.
For DS decoding with phenomenological noise parameters $p$ and $q$, we
used
$$
\Pr(X)=\Pr(Y)=\Pr(Z)=p/3 \mbox{ and } \Pr(\epsilon_j=1)=q.
$$
In addition, one can use several suboptimal decoders with
significantly smaller complexity.  In particular, one may use the
following {\em 2 step algorithm}
 \begin{enumerate}
 \item Construct the syndrome trellis, say $T$, for the DS code with
   $\bG_{\rm DS} = (\bG| \bI_r)$. It will have much smaller number of
   states compared with the trellis for Eq.~(\ref{eq:H-DS}).

 \item Decode $\by$ by the Viterbi decoding of the code with generator
   $(\bI_{n-k},\bA^T)$, to get a tentative syndrome
   $\widehat{\bs}=(\widehat{s}_1,\ldots,\widehat{s}_{n-k})$.
   Typically $\widehat{\bs}$ would have significantly smaller number
   of errors than the measured syndrome.
 \item Decode $({\bf 0},\widehat{\bs})$ by the Viterbi decoding using trellis $T$.
 \end{enumerate}
 Several variations of this algorithm are possible. For example we may
 decode $\by$ using a list decoding of size $L$, get several tentative
 syndromes $\widehat{\bs}_i,i=1,\ldots,L$, and use them in turn in
 step 3 of the above algorithm, and choose the best result.

 Another possibility is to use BCJR decoding for computing the
 tentative syndrome
 $\widehat{\bs}=(\widehat{s}_1,\ldots,\widehat{s}_{n-k})$.

\section{Numerical results} 

We constructed the trellises and numerically analyzed the performance
of several quantum convolutional DS codes differing by the structure
of the binary generating matrix $\bA$.  In all cases, we used as the
starting code the code $\bQ_6$ with parameters $[[24,6,3]]$
constructed from a linear $\mathbb{F}_4$ convolutional code with
generator $\bg=(111|1\omega\bar\omega)$, one of the many QCCs
constructed in Ref.~\cite{Forney-Grassl-Guha-2007}.  As discussed in
Sec.~\ref{sec:background}, the stabilizer generators for codes in this
family have weights $\wt(\bg_j)\in\{3, 6\}$, see
Eq.~(\ref{eq:QCC-G0}).

Specifically, we used the following choices. (\textbf{i}) Code ``GA'',
a quantum DS CC (\ref{eq:G-DS}) with the $16\times 18$ matrix $\bA^T$
chosen as the generating matrix of the binary convolutional code (CC)
with the generator row $\bg=(11|01|11)$.  Explicitly,
{\small\begin{equation}
  \label{eq:AT-GA}
  \bA_{\rm GA}^T=\left(
    \begin{array}[c]{cccccc}
      1\;1&0\;1&1\;1 \\
        &1\;1&0\;1&1\;1 \\ & & \ldots & \ldots &\ldots
    \end{array}\right).
\end{equation}}%
Matrix $\F=\bA_{\rm GA}\bG({\cal Q}_k)$ has row weights
$\wt(\f_j)\in\{6,9\}$.

\noindent(\textbf{ii}) Code ``GR'' (here R stands for ``repetition'')
is constructed similarly, except the matrix $\bA^T$ is formed by a
trivial CC code with $\bg=(11)$.  Explicitly, it has the form
{\small\begin{equation}
  \label{eq:AT-GR}
  \bA_{\rm GR}^T=\left(
    \begin{array}[c]{cccc}
      1\;1& \\
        &1\;1& \\ & & \ldots
    \end{array}\right).
\end{equation}}%
It is easy to see that such a matrix results from three-times repeated
measurement of the original set of generators in the $18$ rows of
$\bG({\cal Q}_6)$.  Respectively, only the original stabilizer
generators of weights $3$ and $6$ need to be measured here.  

\noindent(\textbf{iii}) Code ``GI'' is a trivial DS code with
$\bA_{\rm GI}={\bf 0}$.  The name is due to the structure of the
matrix (\ref{eq:G-DS}): in this case it has the form
$\bG_{\rm DS}=(\bG({\cal Q}_6)\,|\,\bI_{18})$.  With phenomenological
error model (A) [Sec.~\ref{sec:error-model}] and three-times repeated
measurement, we use this code as a simpler alternative to code ``GR''.
Namely, we first perform majority vote on every bit of the syndrome,
then use the DS code GI for actual decoding.  

\noindent(\textbf{iv}) Finally, the code ``G'' stands for yet another
simple DS decoding protocol for three-time repeated measurements.
Again, the syndrome bits are obtained using majority vote, but the
resulting syndrome is considered as error-free, and the decoding is
done directly using the QCC ${\cal Q}_6$.  Main difference with the
previous case is that here a single-bit syndrome error after majority
vote necessarily results in a decoding fault.

\begin{figure}[htbp]
{\includegraphics[width=0.49\columnwidth,clip,trim=2.45in
    4.1in 2.7in 4.25in]
  {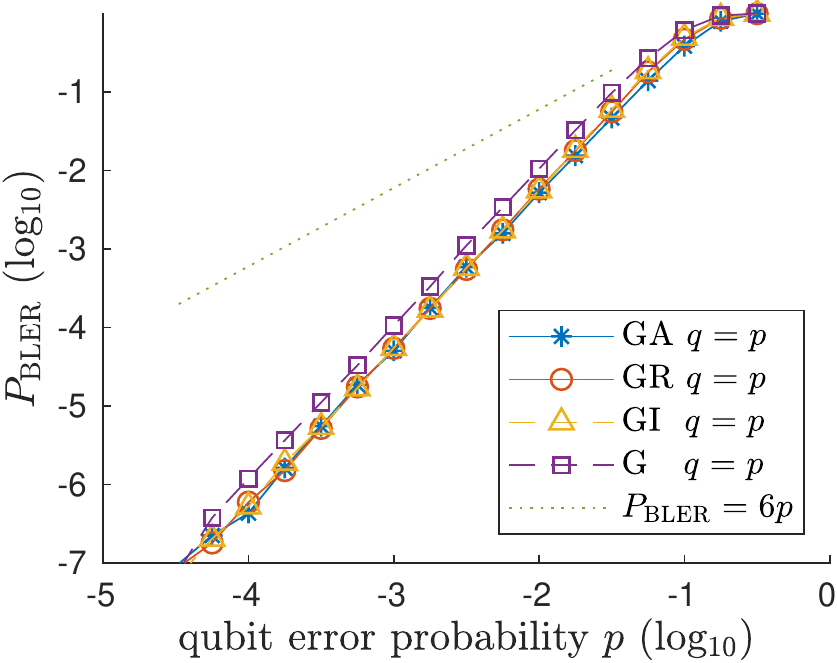}} %
{\includegraphics[width=0.49\columnwidth,clip,trim=2.45in
    4.1in 2.7in 4.25in]
  {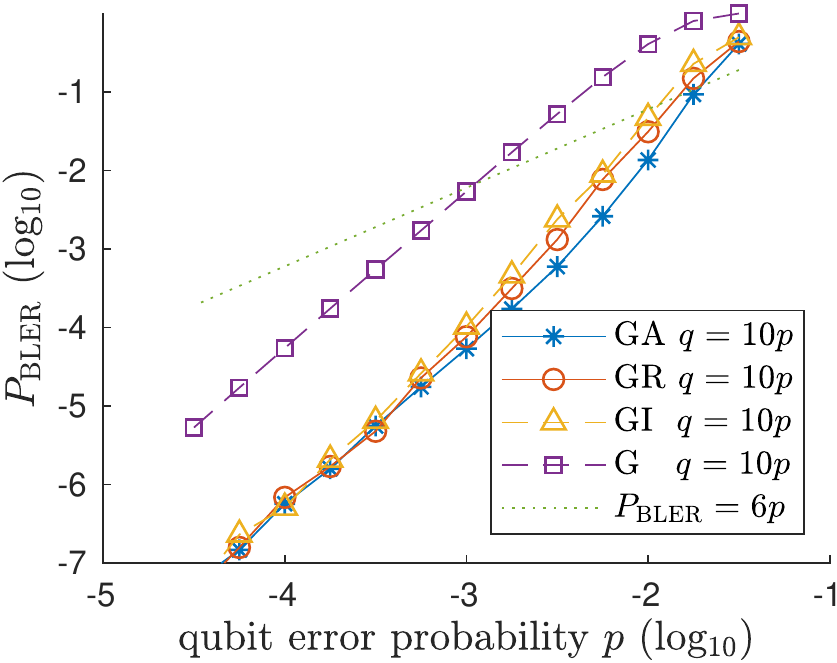}}
\caption{Phenomenological noise model (A) with depolarizing errors
  (probability $p$) and syndrome bit measurement error probability
  $q=p$ (left) and $q=10p$ (right).  Symbols show the block error
  probability $P_{\rm BLER}$ for four decoders as indicated, see text
  for details.  Dotted
  lines give the nominal single-qubit break-even threshold,
  $P_{\rm BLER}=p$.}
  \label{fig:A-four-decoder-q=p}
\end{figure}

Results of simulations with phenomenological error model (A) are shown
in Fig.~\ref{fig:A-four-decoder-q=p}, along with a break-even line
$P_{\rm BLER}=6p$ ($k=6$ unprotected qubits).  We did not attempt to
account for larger weight of measured operators in the case of code
GA.  Single-shot block error probabilities $P_{\rm BLER}$ for four
decoders as indicated in the caption are shown.  For each point,
simulations were done until $N=100$ decoder failures.  The slope is
consistent with the distance $d=3$ of the quantum code.  Results
indicate that (with the exception of the simplest decoder G) all
decoders are able to correct most syndrome measurement errors with
$q=p$, and also for $q=10p$ in the interval $p\lesssim 10^{-3}$.  With
larger error rates, code GA works best, consistent with its larger
distance for syndrome errors.

\begin{figure}[htbp]
  \centering
{\includegraphics[width=0.65\columnwidth,clip,%
    trim=2.45in 4.1in 2.7in 4.25in]
    {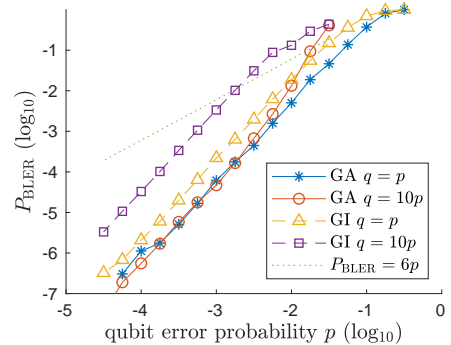}}
\caption{Effective block error rate $P_{\rm BLER}$ with
  phenomenological error model B.  Only results for GA and
  single-interval GI decoders as indicated are shown.}
  \label{fig:B-four-decoder-q=10p}
\end{figure}

In simulations with phenomenological error model (B) we measured the
average fail time of the code\cite{Wang-Fowler-Hollenberg-2011}.
Namely, in each simulation round $j$ repeated decoding cycles are done
until decoding failure after round $t_j$; the corresponding average
after $N\ge 100$ simulation rounds was recorded.  Effective block
error rate $P_{\rm BLER}=1/(\bar t-1)$ was then extracted from the
average fail time $\bar t$ assuming Poisson distribution of life times
$t_j'=t_j-1$ with parameter $\lambda=P_{\rm BLER}$.  We decoded every
cycle $t$ separately, using the accumulated data error
$\hat{\e}_1+\hat{\e}_2+\ldots+\hat{\e}_{t-1}$ found in the previous
cycles as an offset.  Consistent with the standard protocol for
quantum LDPC codes\cite{Wang-Fowler-Hollenberg-2011}, a failure would
be recorded if at time step $t$ decoding with zero syndrome error
$\boldsymbol\epsilon_t={\bf 0}$ gives a logical error.  Otherwise, a
new estimated error $\hat{\e}_t$ would be computed with the syndrome
error $\boldsymbol\epsilon_t$ present, and calculation repeated at
$t=t+1$.  The results are shown in Fig.~\ref{fig:B-four-decoder-q=10p};
they are largely consistent with those for phenomenological error
model (A).

In simulations with circuit error model (C) we constructed the actual
circuits for measuring quantum operators corresponding to rows of
$\bG$, including the redundant rows for code GA, with the attempt to
maximally parallelize the measurements.  We then used a separate
program to generate random Pauli errors with probability $p_1$ per
interval between the gates, propagated the errors through the circuit,
and recorded the actual accumulated error $\underline\e_t$ and the
measured syndrome $\bs_t$ at the end of each measurement cycle
$t=1,2,\ldots$.  Additional syndrome measurement error $q_1$ was added
at the time of subsequent processing.  These data then have been used
with the decoders identical to those for model B.

\begin{figure}[htbp]
  \centering
  {\includegraphics[width=0.68\columnwidth,clip,%
    trim=2.45in 4.1in 2.6in 4.25in]
{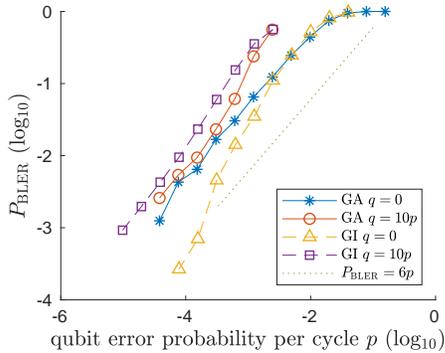}}
\caption{Effective block error rate (per gate) for phenomenological
  error model C as a function of gate error probability $p_1$ scaled
  by cycle duration.}
  \label{fig:C-four-decoder-q=10p}
\end{figure}

The obtained effective block error rates are plotted in
Fig.~\ref{fig:C-four-decoder-q=10p}.  One striking difference with
phenomenological error models A and B is that the calculated curves no
longer have quadratic dependence on BER, as would be expected for a
code with distance $d=3$.  The reason is that we have used non-FT
circuits in simulations.  As a result, e.g., a single ancilla error
can propagate and multiply through the circuit, resulting in a
higher-weight error which cannot be corrected by the code.

\section{Discussion and Future Work} 

In conclusion, in this work we introduced quantum convolutional
data-syndrome codes, constructed an efficient decoder for this class
of codes, and analyzed numerically the performance of a family of DS
codes based on a single QCC with parameters $[[24,6,3]]$ using three
distinct error models.  In particular, this was the first time a DS
code has been simulated with the circuit error model.  

Here we exclusively relied on the QCCs designed in
Ref.~\cite{Forney-Grassl-Guha-2007}.  These codes have relatively high
weights of stabilizer generators.  It is an open question whether
degenerate QCCs exist, with small-weight generators, large distances,
and trellises with reasonably small memory sizes.  For the
purpose of constructing convolutional DS codes, one would further like
to have a QCC with a redundant set of minimum-weight stabilizer
generators.  For such codes, degenerate Viterbi decoding
algorithm\cite{Pelchat-Poulin-2013} would be particularly useful.

Our limited simulation results indicate that a DS code with
large-distance classical syndrome code may show competitive
performance in the regime where measurement errors are significant,
even though the corresponding generators may have larger weights.
This regime is experimentally relevant, e.g., for superconducting
transmon qubits with dispersive readout, where measurement time can be
as large as 500ns, compared to under 50ns two-qubit gates, with the
error probabilities scaling accordingly.  It is an open question
whether similarly constructed non-convolutional DS codes could be
useful in this regime, e.g., for optimizing the performance of surface
codes in the current or near-future generation of quantum computers.

One obvious way to improve the practical performance of DS codes is by
using FT gadgets for generator measurements, to control error
propagation.  In particular, we intend to try flag measurement
circuits\cite{Chamberland-Beverland-2017}, as this technique has
relatively small overhead in the number of qubits.


\end{document}